# Viewpoint: Can Symmetric Tilt Grain Boundaries Represent Polycrystals?

Malik Wagih and Christopher A. Schuh[1]

*Department of Materials Science and Engineering, Massachusetts Institute of Technology,
77 Massachusetts Avenue, Cambridge, MA 02139, USA*

**Abstract**

Grain boundaries control a wide variety of bulk properties in polycrystalline materials, so simulation methods like density functional theory are routinely used to study their structure-property relationships. A standard practice for such simulations is to use compact, high-symmetry (coincident site lattice) boundaries as representatives of the much more complex polycrystalline grain boundaries. In this letter, we question this practice by quantitatively comparing the spectra of atomic sites and properties amongst grain boundaries. We show, using solute segregation as an example property, that highly symmetric tilt boundaries (with $\Sigma$ values less than 10) will fail to capture polycrystalline grain boundary environments, and thus lead to incorrect quantitative and qualitative insights into their behavior.

Grain boundaries (GBs) have a tremendous influence on the bulk structural and functional properties of polycrystalline materials [1]. GBs generally have a structurally disordered nature with a wide variety of site-types or local atomic environments (LAEs) that differ from the bulk lattice environment. The properties of a GB, as well as the nature of its interactions with other defects in the material, are determined by the array of LAEs present at the boundary. Therefore, since the advent of atomistic simulations, tremendous effort has been made to probe GBs at the atomic level to understand their structure-chemistry-property relationships. This includes, but is not limited to, the study of GB cohesion [2,3], diffusion [4,5], mobility [6,7], sliding [8,9], fracture [10], irradiation resistance [11], and phase transformations [12], as well as interactions between GBs and dislocations [13–15], vacancies [16], and solutes [17,18]. Simulations of this kind provide invaluable insights into GB behavior and are indispensable tools for understanding GB-driven phenomena in materials.

To date, most atomistic studies on GBs focus on coincident site lattice GBs [19–22], which are characterized by their reciprocal density of coincident sites $\Sigma$, and almost exclusively explore symmetric tilt grain boundaries (STGBs) of high coincidence (i.e., high symmetry) with $\Sigma$ values that are typically less than 10. STGBs are chosen because they provide compact model GBs that can be constructed with small bi-crystals of O(100) atoms (i.e., on the order of 100 atoms)—this makes them accessible to almost all atomistic simulation methods including density functional theory, which is typically limited to at most O(1000) atoms. For example, the Crystalium database [23] (part of the Materials Project [24]) for elemental grain boundaries is currently limited to <$\Sigma$10 STGBs, which is typical of most quantum-accurate studies. In fact, the largest fraction of atomistic studies focus on just two STGBs: $\Sigma5(021)$ and $\Sigma5(013)$; a good illustration of this can be found in the review paper by Lejček et al. [25] on solute segregation in GBs, which shows that over 70% of surveyed atomistic studies in the literature study the phenomenon using those two $\Sigma5$ STGBs. These are the smallest nontrivial GB structures that can be simulated, but their popularity for modeling work defies the experimental fact that they are seldom sampled in polycrystalline GB networks [26,27], as shown for example, by the extensive experimental work by Rohrer and co-workers in aluminum [28] and nickel [29]. Besides Σ5, it is important to also note that even when other low-Σ tilt boundaries are observed in polycrystals [28–30], they are often not symmetric. Observations of this type even led Merkle and Wolf [30] to suggest, over thirty years ago, that asymmetric boundaries should be generally more favorable than symmetric ones in polycrystals, since they often have lower energies and more degenerate structures (with similar energies) [30,31].

---

[1] Corresponding author. Email address: schuh@mit.edu (Christopher A. Schuh).

The above concern speaks to a deeper philosophical question that we pose in this viewpoint article: Can high-symmetry STGBs represent (or approximate) the GB network in polycrystals? Can narrow simulation studies on STGBs provide quantitative or qualitative insights into GB-driven phenomena that will have relevance outside of their highly idealized conditions? Although the tacit assumption of the GB modeling community is, at least from a qualitative perspective, "yes", here we challenge this assumption and explore the limitations of STGBs as proxies for true GB structure and behavior. We focus on STGBs with low Σ values (< Σ100) that can be constructed using bi-crystals with O(1000) atoms, which makes them, in principle, accessible to quantum-accurate simulation methods. We use GB solute segregation as an example to illustrate the limitations of STGBs, but our findings should be extensible to other GB-driven phenomena.

Since GB structure begets GB properties, the appropriate starting point to address this problem is with an understanding of the range of local structures, or LAEs, at STGBs vis-à-vis the grain boundary network in polycrystals. Such a comparison is timely, as it is only recently that full populations of LAEs have been explored for polycrystals, and there are also now datasets connecting LAE structure to the property of GB segregation [32–34]. We therefore address the question of how STGBs compare with polycrystals directly through computation of structure and segregation properties. We use Ag as a model solvent, and explore the GB segregation of various solute elements in the dilute limit. (See supplementary section SM.1 for more details on the computational methods [32,35–52]).

In Fig. 1(a), we show the 20x20x20 nm$^3$ Ag polycrystal considered in this work, which has 16 grains with an average grain size of ~10 nm, and ~10$^5$ GB sites for a total GB atomic site fraction of ~22%. Previously, in Ref. [53], we have shown that a polycrystal of this size will result in a similar distribution of segregation energies to that of a larger polycrystal with hundreds of grain boundaries that randomly and nearly-uniformly sample the GB disorientation space (as well as that of multiple other polycrystals); this suggests that the (20 nm)$^3$ polycrystal provides a good sampling of the finite variety of LAEs that exist in the polycrystalline GB space. Even if one were to argue that the polycrystal explored here covers only a small subset of the GB LAE space, it nonetheless provides a good test case for the ability of STGBs to capture the polycrystalline space, since it should be, in principle, easier to capture than the entire GB LAE space.

To characterize the nature of GB LAEs sampled in the Ag polycrystal, we borrow from the literature on machine-learned interatomic potentials [54,55], which develops complex methods for encoding the LAE of an atomic site within a cutoff distance into a feature vector that preserves the material physical symmetries (e.g., permutation, rotation, translation, etc.). Here, we employ the smooth overlap of atomic positions (SOAP) [43,44] descriptor to capture the LAE of the Ag GB sites up to and including third nearest neighbor atoms (a cutoff distance of 5.5 Å). Since the neighboring atoms influence the response of a GB site to external stimuli, including them in the LAE is crucial. And thus, the SOAP descriptor offers a much richer representation of the GB LAEs than is possible with simple classical descriptors, such as atomic volume and coordination, which can only describe the LAE of a GB site up to (but excluding) its first nearest neighbors. (See Refs. [33,34,56] for further information about using SOAP to represent LAEs in GBs and to predict solute segregation. And, see supplementary section SM.2 for details about calculating the descriptors used here).

To visualize the multidimensional SOAP vector (1,225 features in this work), we project it into a low-dimensional (reduced) space using principal component analysis [57,58] (i.e., we compress the SOAP vector into a few principal components that account for most of the variance in the data). Fig. 1(b) shows, for the Ag polycrystal, the values for the first two principal components, which capture ~80% of the variance in the system. The vertical and horizontal lines represent the values for a Ag bulk lattice site; interestingly and not surprisingly, bulk sites are an extreme edge-case for LAE amongst the rich diversity of GB sites.

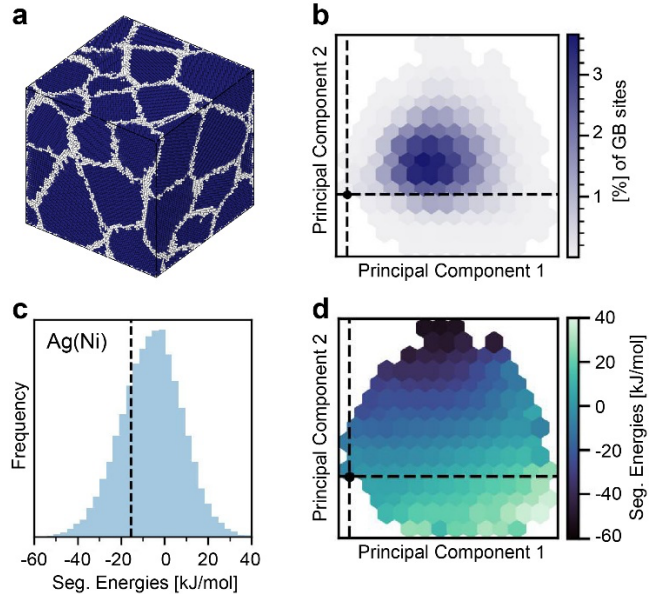

Fig. 1: (a) 20x20x20 nm$^3$ Ag polycrystal that has 16 grains of ~10 nm size; its ~10$^5$ GB atoms are highlighted in light gray. (b) The density of GB LAEs across the reduced (by principal component analysis) SOAP-descriptor space; the dashed lines intersect at the values of a crystalline bulk site. (c) The spectrum of dilute-limit segregation energies for a Ni solute at all GB sites; the dashed vertical line is the first quartile value. (d) The variation of Ni segregation energetics across the reduced SOAP-descriptor space.

With a detailed quantitative view of the LAE spectrum in the polycrystal, we turn in Fig. 1(c) to an associated property spectrum: the segregation energies of a Ni solute. The distribution is close to normal, as expected for polycrystals [33,34,53,59–61], and it maps rather smoothly to the structure of the individual sites as seen in Fig. 1(d) in the reduced SOAP space; the most attractive GB sites for Ni segregation fall in the top of Fig. 1(d). It is interesting to note that the most favorable GB sites are those that are most distant from (i.e., least similar to) the Ag bulk lattice LAE (dashed lines in Fig. 1(d)), and not those most frequently sampled in the polycrystal (the dark blue region in Fig. 1(b)).

In this letter, we will study three sets of STGBs (see supplementary Table S1 for more details on the 60 STGBs studied here):

1) **{Σ5}**: Σ5(021) and Σ5(013). As noted above, these two STGBs are the subject of most atomistic studies in the literature.
2) **{<Σ10}**: a set of 8 STGBs that are accessible to quantum-accurate methods, e.g., density functional theory, and as such, are often examined in the literature.
3) **{<Σ100}**: 60 STGBs that include lower-symmetry cases that are technically accessible, but rarely, if ever, studied by quantum-accurate methods because they approach computational limits.

We will focus our attention first on the {Σ5} STGBs visualized in Fig. 2(a) and (b), in light of their outsized prevalence in atomistic studies. In Fig. 2(c), we show where the seven unique GB sites for the Σ5(021) and Σ5(013) STGBs are located in the reduced SOAP space. In this representation we now limit the color map to the most physically-important top-quartile value for Ni segregation (segregation energy less than -15 kJ/mol, shown by the dashed line in Fig. 1(c)). The presentation in Fig. 2(c) highlights the subset of GB space that constitutes the most attractive GB sites for segregation, and controls the extent of segregation in the alloy. This manner of presentation makes it very clear that the {Σ5} STGBs simply do not sample the corner of LAE space where Ni segregation is most attractive. In fact, only three sites even approach the top-quartile of segregation energy that is most physically relevant, as seen in Fig. 2(d). This poor sampling of the true spectrum is expected to severely impact quantitative predictions of segregation, which can be made with the spectral isotherm [53,62,63]:

$$X^{tot} = (1 - f^{gb}) \cdot X^c + f^{gb} \cdot \sum_{\Delta E_i^{seg} = -\infty}^{\infty} n_i^{gb}(\Delta E_i^{seg}) \cdot \left[1 + \frac{1 - X^c}{X^c} \cdot \exp\left(\frac{\Delta E_i^{seg}}{k_B T}\right)\right]^{-1} d(\Delta E_i^{seg}) \quad (1)$$

where $X^{tot}$ is the total solute concentration; $X^c$ is the equilibrium bulk solute concentration; $f^{gb}$ is the GB atomic site fraction, which can be approximated by $f^{gb} = 1 - [(d - t)/d]^3$ where (d) and (t) are the grain size and GB thickness, respectively; $n_i^{gb}(\Delta E_i^{seg})$ is the discrete distribution function for segregation energies at the GB; $k_B$ is Boltzmann's constant; and T is temperature. The equilibrium average solute concentration at the GB, $\overline{X}^{gb}$, is simply the right-hand summation in Eq. (1), i.e., $\overline{X}^{gb} = [X^{tot} = (1 - f^{gb}) \cdot X^c] / f^{gb}$.

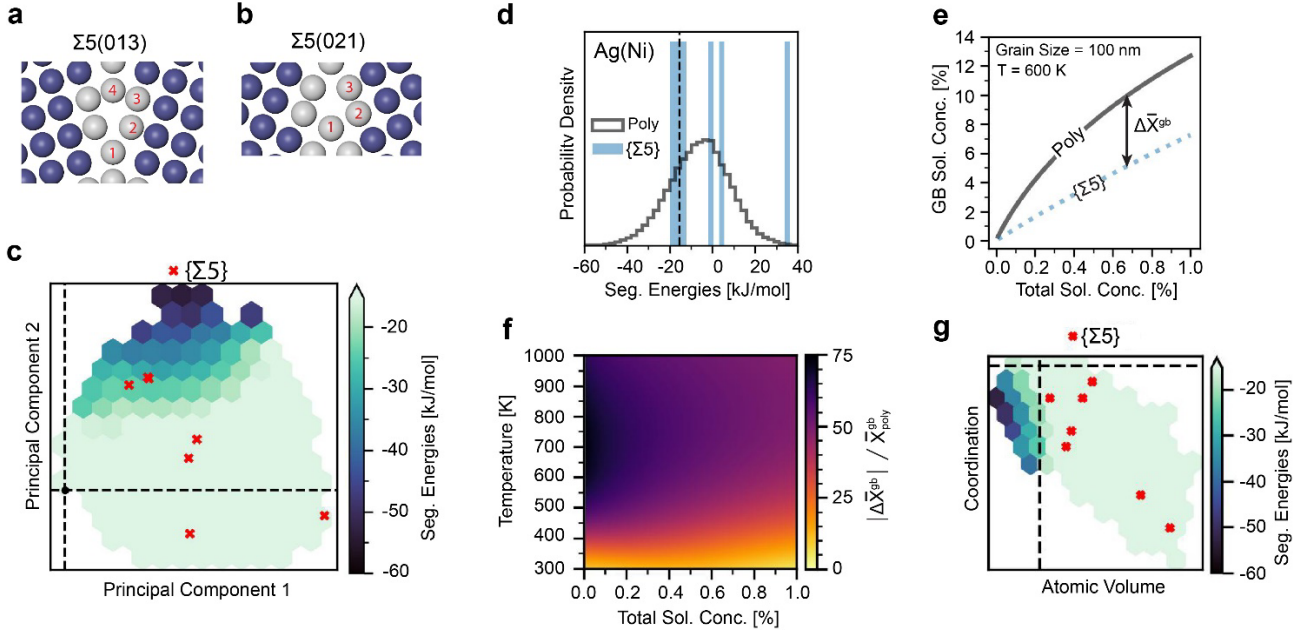

Fig. 2: A visualization of the GB sites for the {Σ5} STGB set, which includes the (a) Σ5(021) and (b) Σ5(013) STGBs. (c) The seven unique LAEs for the {Σ5} are mapped across the reduced SOAP-descriptor space for the Ag polycrystal in Fig. 1; the color map is limited to a maximum of -15 kJ/mol to highlight the most attractive tail for Ni segregation spectrum as shown in Fig. 1(c). (d) A comparison of the discrete distributions, $n_i^{gb}(\Delta E_i^{seg})$, obtained for Ni segregation in the polycrystal and the {Σ5}. (e) The predicted equilibrium segregation state, using Eq. (1), for a 100 nm polycrystal at T=600 K using the different spectra from (d); the difference in predictions for the solute content at the GB is $\Delta \overline{X}^{gb} = \overline{X}^{gb}_{stgb} - \overline{X}^{gb}_{poly}$. (f) The error in predictions for $\overline{X}^{gb}$ across the composition-temperature space. (g) Recasting the SOAP-based map in (c) in terms of two classical descriptors for the LAE: atomic volume and coordination.

In Fig. 2(e), we compare the predicted equilibrium segregation state for a Ag polycrystal with a grain size of 100 nm and a temperature of 600 K, using the two different distribution functions $n_i^{gb}(\Delta E_i^{seg})$ for the {Σ5} STGBs and the polycrystal from Fig. 2(d). As expected, the {Σ5} STGBs severely underpredict (by more than half) the equilibrium solute content in the polycrystalline GB network (even for relatively dilute solute concentrations of $X^{tot} < 1\%$). We further explore such quantitative failures in Fig. 2(f), where we plot the error in GB solute content at (defined as $\varepsilon^{gb} = \left|\overline{X}_{stgb}^{gb} - \overline{X}_{poly}^{gb}\right|/\overline{X}_{poly}^{gb}$, where $\overline{X}_{stgb}^{gb}$ and $\overline{X}_{poly}^{gb}$ are the calculated solute concentrations using Eq. (1)). The {Σ5} STGBs fail to predict, by a large magnitude, the correct segregation behavior across the composition–temperature space, with the error ($\varepsilon^{gb}$) exceeding 60% for a sizable subset of the space.

In some sense, the failure of {Σ5} STGBs to capture physically reasonable segregation behavior should be expected, as those boundaries are highly symmetric and coherent, whereas the most potent sites for segregation are generally incoherent and asymmetric. This is entirely a sampling issue: the failure of STGBs is not due to their inability to replicate polycrystalline LAEs, i.e., a bi-crystal vs. polycrystal simulation setup issue, but rather due to their failure to sample enough relevant LAEs. STGBs are, in fact, quite capable of mimicking polycrystalline LAEs and their response to stimuli, as illustrated in supplementary Fig. S2, which shows that the seven unique sites of the {Σ5} STGBs produce similar segregation energies to those of the closest polycrystalline sites in SOAP-descriptor space (i.e., environments most similar to the STGB sites).

The sampling failure of {Σ5} STGBs can be seen more explicitly by taking a different, more classical view of the LAE. In Fig. 2(g), we recast all of the GB sites in the polycrystal from Fig. 2(c); instead of the complex SOAP-based LAE, here we use a near-neighbor-based view on axes of local atomic volume of the site (Voronoi volume) and coordination number (number of neighbors). These are classic atomic environment descriptors generally expected to correlate with segregation behavior, on the grounds that in the current case where the solute is smaller than the host, more coordinated locations with tight atomic environments offer geometrically (and thus energetically) better sites for the solutes. Fig. 2(g) bears this expectation out for the full spectral view of GB sites, and it even more clearly illustrates the failure of the {Σ5} STGB sites to sample the relevant tail of the distribution.

Despite the above quantitative failures of {Σ5} STGBs as model boundaries, one might be tempted to argue that their value lies primarily in gleaning *qualitative* predictions, such as screening for alloying elements that are more or less likely to segregate, rather than precise quantitative predictions. Unfortunately, though, we find that {Σ5} STGBs also fail to provide the correct qualitative trends across the alloy space. Fig. 3(a) shows the computed segregation spectra for four other solutes, Au, Cu, Pd, and Pt in the same Ag polycrystal (Fig. 1(a)), against that of the seven unique sites of the {Σ5} STGBs. Similarly to the earlier case of Ni, the {Σ5} STGBs fail to capture, to different degrees, the most attractive sites of the polycrystalline spectra for all of these solutes. And by extension, the {Σ5} STGBs fail to even correctly order the segregation tendency of these solutes: as seen in Fig. 3(b), for a grain size of 100 nm, $X^{tot}=1\%$, and T=600 K, the {Σ5} STGBs rank the segregation strength of the five solutes as Pt < Pd < Au < Cu < Ni. The correct order from the full polycrystal models is Au < Pd < Pt < Cu < Ni, requiring at least two pair-wise swaps to correct the faulty sequence.

To quantify such ranking errors, we use the normalized Kendall tau rank distance [64], $\tau_K$, which counts the number of pairwise disagreements between two lists, and has a value in the interval [0,1], for perfect ranking (0) to perfectly inverse/incorrect ranking (1). In this case, $\tau_K$ has a very high value of 0.3, which means 30% of pairwise comparisons differ in order from the correct ones (as predicted by the polycrystal). Such disagreement persists over large regions of composition and temperature space, as mapped explicitly in Fig. 3(c). While the $\tau_K$ error is not constant, i.e., the accuracy of qualitative predictions is not consistent across the composition–temperature space, it is clear that in general, the STGBs fail to capture the correct physical segregation and even the relative ranking between different alloys at essentially all compositions and temperatures.

We would like to highlight two points here. First, the choice of interatomic potential should not greatly affect the results. The limitations of STGBs arise from their failure to sample the GB sites of a polycrystal in a structural (geometric) sense as shown in Fig. 2(c) and (g). This holds true regardless of how this failure may impact GB properties (such as segregation energies) which are sensitive to the interatomic potential. This point is tested in supplementary Fig. S3, in which the results obtained in Fig. 2 are reproduced using a more sophisticated embedded atom potential for Ag-Ni developed by Pan et al. [51] that is validated against density functional theory calculations for GB segregation. Although the magnitudes of segregation energies differ slightly from those of the Adams et al. [47] potential we use in the body of this letter, our qualitative and quantitative conclusions remain the same. More broadly, the present results should not be construed as somehow limited to Ag or Ag alloys; any metallic potential that produces an FCC packing would be expected to generate similar GB structures, and the STGBs would still exhibit the same sampling concern relative to a polycrystal. While the subset of physically-important LAEs in a polycrystal changes depending on the base metal, polycrystal texture, and GB property of interest, the question of how STGBs sample that subset remains. Second, the difference between polycrystals and STGBs is not solely due to triple junctions (although triple junctions play an important role in polycrystals, especially at nanocrystalline grain sizes [65,66]). This would be true if triple junctions were responsible for the most attractive GB sites for segregation. However, as can be seen in supplementary Fig. S4, the most attractive sites for segregation are not confined to triple junctions and are well spread throughout the polycrystal.

The above findings argue that the {Σ5} STGBs should not be used to model GB segregation in polycrystals, as they fail both quantitatively and for qualitative screening. Whether this kind of sampling failure matters for other situations and properties is worthy of further study. However, it remains an open question whether or not adding more STGBs to the {Σ5} population can improve the sampling to the point of physical relevance in the polycrystal case. We explore this issue in full detail for the {<Σ10} and {<Σ100} STGB sets next.

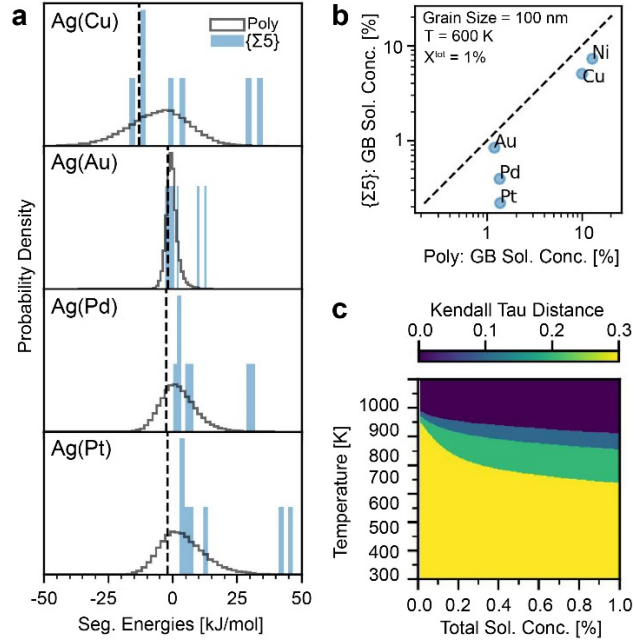

Fig. 3: (a) The spectrum of segregation energies for a solute of Cu, Au, Pd, and Pt in the Ag polycrystal in Fig. 1(a); the dashed vertical lines are the first quartile values for each spectrum. (b) A comparison of the predictions of equilibrium $\overline{X}^{gb}$ for five solutes using the computed spectra for the {Σ5} against that of the polycrystal, as shown in (a) for four solutes, and in Fig. 2(d) for Ni. The error in ranking the five solutes in terms of segregation strength using the {Σ5} spectra as quantified by the Kendall tau rank distance.

Let us first consider the set {<Σ10}, which consists of eight STGBs that are most frequently simulated with quantum-accurate methods (including the two {Σ5} STGBs). Fig. 4(a) shows that, although the ~40 unique LAEs of the {<Σ10} allow for a slightly better sampling of the polycrystalline LAE space, they still do not entirely capture it, nor do they properly sample the most important regions within it. (Again, this is supported by the simple LAE descriptors of atomic volume and coordination as shown in supplementary Fig. S5(a)). Therefore, similar to the {Σ5} set, the {<Σ10} STGBs provide fundamentally different LAE statistics than the polycrystal, as shown in Fig. 4(b). This leads to inaccurate physical insights into the composition-temperature dependent behavior of solute segregation in the polycrystal, both quantitively and qualitatively, as shown in Fig. 4(c) and Fig. 4(d), respectively.

The situation improves remarkably for the {<Σ100} STGBs. As shown in Fig. 4(e) and Fig. 4(f), the {<Σ100} STGBs provide a noticeable improvement in the degree of sampling of the LAE space, which leads to a significant improvement in the accuracy of quantitative predictions across the composition-temperature space, as shown in Fig. 4(g), with the error being less than 10% for most conditions. The qualitative results, however, still show errors in ranking the solutes across portions of the composition-temperature space (primarily at high temperatures), as shown in Fig. 4(h). This suggests that including higher Σ values in simulations of GB structure and properties may be a viable path to properly sample the true range of polycrystalline environments, which agrees with experimental efforts in this vein [67]. Unfortunately it is extremely rare to simulate such high Σ values, and inspection of Fig. 4(e) suggests that values above Σ100 may be needed for full coverage of the LAE space. Importantly, the statistics of sampling in the polycrystalline environment are certainly not mirrored in any given collection of STGBs, so more work is needed to develop a proper selection of LAEs across different STGBs to mimic the statistics of the polycrystal (e.g., fewer high-volume LAEs). Perhaps the largest challenge to the STGB approach, however, is that already many of the {<Σ100} boundaries are almost prohibitively expensive for quantum-accurate methods, such as density functional theory, especially for high-throughput material discovery and design. This argues for increased use of multiscale methods [68–71] for these problems, which, although only rarely used for GB problems [34,41,72], do permit sampling of full polycrystalline GB environments [34]. Lastly, this work also speaks to the importance of simulation efforts [73–78] that go beyond STGBs to gain a comprehensive understanding of GB properties across the complex GB five-dimensional space.

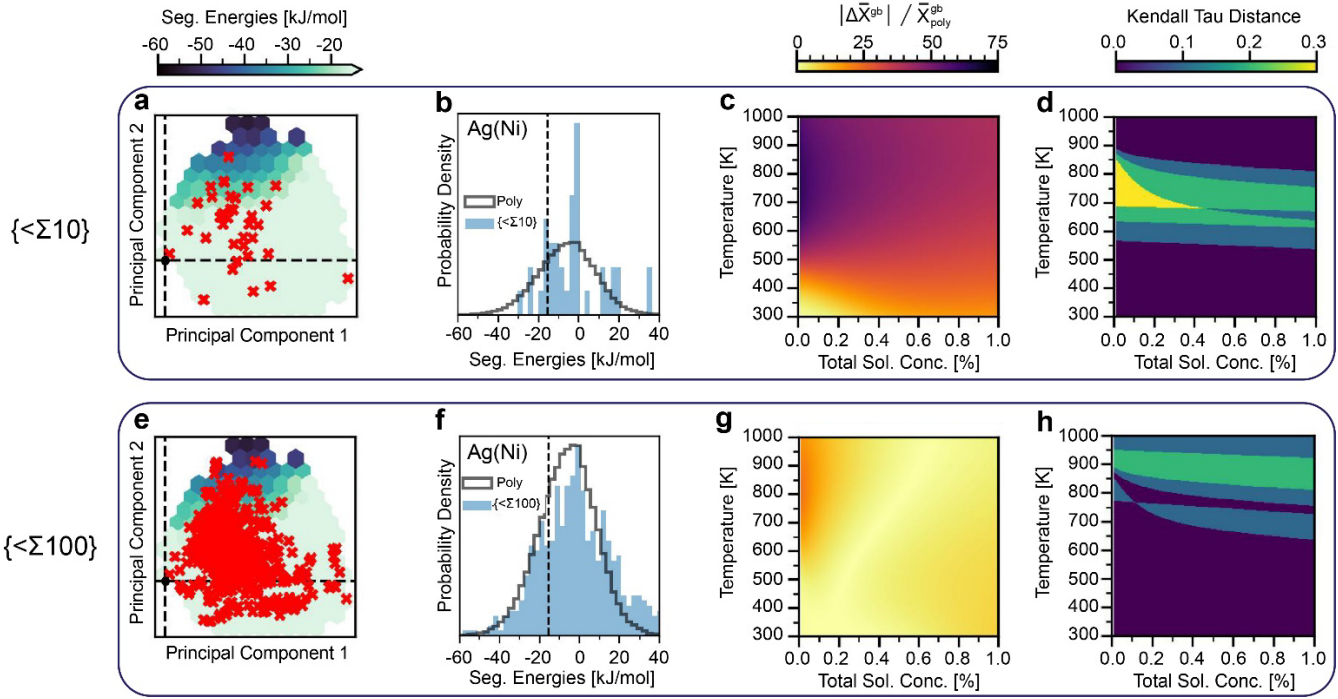

Fig. 4: For the {<Σ10} and {<Σ100} STGB sets, respectively, we show their coverage of polycrystalline LAE space in (a) and (e); their computed spectra for Ni segregation against that of the polycrystal in (b) and (f); their quantitative error in predicting the equilibrium solute content at the GB for Ag(Ni) in (c) and (g); and their qualitative error, quantified by $\tau_K$, in ranking the solutes Au, Cu, Ni, Pd, and Pt in terms of segregation strength in (d) and (h).

In conclusion, we have shown the pitfalls of using a small subset of STGBs as representative GBs to study the structure-property relationships for grain boundaries in general. We have demonstrated that the high-symmetry STGBs typically used in atomistic studies, particularly of the quantum-accurate type, do not capture polycrystalline GB environments, and can lead to incorrect quantitative and qualitative insights into their behavior, as shown for the case of GB solute segregation. It is only with STGBs with lower symmetry (as we increase the Σ values from 10 to 100) that we begin to achieve reasonable coverage of the atomic environment space; however, many of these STGBs are prohibitively expensive for quantum-accurate simulation. It is thus ever more important to develop tools that can directly probe polycrystalline GB environments, such as quantum-accurate interatomic potentials, and multiscale simulation techniques. We hope this viewpoint will inspire the materials simulation community to develop new approaches to GB-related problems generally.

## Acknowledgments

This work was supported by the US Department of Energy, Office of Basic Energy Sciences under grant number DE-SC0020180.

# SUPPLEMENTAL MATERIAL

# Can Symmetric Tilt Grain Boundaries Represent Polycrystals?

Malik Wagih and Christopher A. Schuh[*]

Department of Materials Science and Engineering, Massachusetts Institute of Technology,
77 Massachusetts Avenue, Cambridge, Massachusetts 02139, USA.

## SM.1 The Ag Polycrystal

LAMMPS [1–3] is used to perform all molecular statics and dynamics simulations. The embedded atom interatomic potential by Adams et al. [4] is used to describe the Ag polycrystal, as well as its interactions with the solutes Au, Cu, Ni, Pd, and Pt. The software OVITO [5] is used for visualization and structural identification

We use the Ag polycrystal generated in our earlier work in Ref. [6]. To build it, a 20x20x20 nm$^3$ Ag polycrystal is generated with 16 randomly oriented grains using Voronoi tessellations with Atomsk [7]. To relax the grain boundaries (GBs), we perform molecular dynamics simulations using the interatomic potential by Pan et al. [8] to thermally anneal the polycrystal at 600 K and zero pressure for 500 ps and a timestep of 1 fs under a Nose-Hoover thermostat/barostat. After cooling to 0 K with a rate of 3 K/ps, a conjugate gradient structural relaxation is performed. This is followed by a final additional conjugate gradient structural relaxation for both atomic positions and box dimensions (i.e., lattice parameter) using the interatomic potential by Adams et al. [4] noted earlier. The relaxed polycrystal has 444,338 atoms of which 97,167 atoms are at the GB, as identified by the adaptive-common neighbor analysis method [9]; all non-fcc atoms are assumed to belong to the GB.

To compute the spectrum of segregation energies at the GB for a solute atom, we follow the procedure outlined in Refs. [10,11]. For every GB site (i), its segregation energy, $\Delta E_i^{seg}$, is the change in the internal energy (enthalpy) of the polycrystal, when a solute atom segregates from a bulk lattice site to the GB site (i):

$$\Delta E_i^{seg} = E_{gb,i=sol}^{poly} - E_{c=sol}^{poly} \qquad (S.1)$$

where $E_{gb,i=sol}^{poly}$, $E_{c=sol}^{poly}$ is the total relaxed energy of the polycrystal with the solute sitting at GB site (i) and a bulk site, respectively. For $E_{c=sol}^{poly}$, we chose an atomic site in the polycrystal that is surrounded by 3 nm of fcc atoms in all directions to avoid any long-range interactions with GBs. We note that, in this notation, a negative $\Delta E_i^{seg}$ signifies a GB site favorable to solute segregation.

## SM.2 Descriptors for the GB local atomic environments

For each GB site, its local atomic environment (LAE) within a radius of 5.5 Å (which is approximately the midpoint between the third and fourth nearest neighbor distances in fcc Ag) is transformed into a feature vector using the smooth overlap of atomic positions (SOAP) method [12,13] using 16 radial basis functions, 8 degrees of spherical harmonics, and 0.5 Å smearing width for the gaussian functions, which results in a SOAP feature vector of 1,225 features. The code QUIP/GAP [12,14,15] is used to generate the SOAP descriptors.

---

[*] Corresponding author. Email address: schuh@mit.edu .

For the classical LAE descriptors of atomic volume and coordination, the Voronoi atomic volume is calculated using OVITO with a relative face area threshold of 1%, and the coordination number ($N_i$) is calculated using the Fermi-smeared version introduced in Refs. [16,17]:

$$N_i = \sum_j \frac{1}{1 + \exp\left(\frac{r_{ij} - r_0}{\sigma}\right)} \tag{S.2}$$

where $r_{ij}$ is the distance between the GB site (i) and its neighbor atom j, and $r_0$ is defined as $r_0 = 0.5(r_{1NN} + r_{2NN})$, and the smearing parameter $\sigma = 0.2(r_{2NN} - r_{1NN})$, where $r_{1NN}$ and $r_{2NN}$ are the first and second nearest neighbor distances in fcc Ag, respectively.

## SM.3 Symmetric tilt grain boundaries

The code package "GB Code" [18] is used to generate the 60 symmetric tilt grain boundaries (STGBs) considered in this work. For each STGB, we generate 100 initial structures corresponding to 100 rigid translations in the GB plane (Y-Z plane) in both directions to explore the microscopic degrees of freedom of the GB. Since the simulation cells are periodic structures, we maintain a minimum distance of 5 nm between the GB and its crystallographically equivalent image to minimize any interactions [19] between the two boundaries. For each of the initial structures, its energy is minimized (relaxing both atomic positions and cell dimensions) using the conjugate gradient algorithm (with an energy convergence criterion of $10^{-25}$), followed by a force minimization using the Fire algorithm [20] (with a force convergence criterion of $10^{-6}$ eV/Å). Finally, we choose the structure with the lowest energy as the representative one for the STGB; if the energy difference between the top-ranked (least energy) structures is less than 0.01 eV/atom, we visually inspect the structures to choose a representative one. In Fig. S1(a), we show as an example, the Σ5(013) STGB generated using this procedure. In Table S1, we list all 60 STGBs along with their atomic size and cell dimensions.

For each STGB, its GB sites are identified similarly to the procedure for the polycrystal, using the adaptive common neighbor analysis method, where all non-fcc atoms are considered GB atoms. To identify the unique sites (or LAEs), we compare their SOAP descriptors. We consider any two GB sites to be similar (non-unique) if the dot product of their normalized SOAP vectors is ≥0.999 (the dot product of a normalized SOAP vector with itself is 1). The method was tested and validated for a few STGBs.

To compute the segregation energies, we first repeat the GB structures along the GB plane directions (Y and Z) such that the distance between a GB site and its periodic image is at least 5 nm. The larger cell dimensions minimize (or even eliminate) the impact of any self-interactions [21] between the solute atom and its periodic image on the computed segregation energies, i.e., reducing the impact of solute-solute interactions. In Fig. S1(b), we show the larger version of the Σ5(013) STGB structure that was used to compute the segregation energies at its four unique GB sites. We calculate the segregation energy for a GB site (i) in the STGB as:

$$\Delta E_i^{seg} = \left(E_{gb,i=sol}^{stgb} - E^{stgb}\right) - \left(E_{sol}^{crystal} - E^{crystal}\right) \tag{S.3}$$

where $E_{gb,i=sol}^{stgb}$, $E^{stgb}$ is the total relaxed energy of the STGB with and without the solute sitting at GB site (i), respectively. $E_{sol}^{crystal}$, $E^{crystal}$ is the total relaxed energy of a large Ag ideal fcc supercell [22] (with a cell size > 10 nm in all directions) with and without a solute atom substitutionally added to the supercell, respectively.

Table S1: A list of the 60 STGBs considered in this work. We selected STGBs with a Σ < 100, and fewer than 1000 atoms. We note though, that the STGB cells used for segregation calculations are larger than the ones listed here, as we repeat the GB structures along the Y and Z directions (until the cell dimension is greater than 5 nm in those directions) to minimize the impact of any self-interactions for the solute atom (with its periodic image).

| STGB | Dimensions (Å) | | | No. of atoms | STGB | Dimensions (Å) | | | No. of atoms |
| --- | --- | --- | --- | --- | --- | --- | --- | --- | --- |
| | X | Y | Z | | | X | Y | Z | |
| Σ3(1 $\bar{1}$ $\bar{2}$) | 130.2 | 2.9 | 7.1 | 156 | Σ39(5 $\bar{7}$ 2) | 144.8 | 7.1 | 10.4 | 622 |
| Σ3(1 $\bar{1}$ 1) | 127.5 | 2.9 | 5.0 | 108 | Σ41(3 $\bar{3}$ $\bar{8}$) | 147.8 | 2.9 | 26.2 | 654 |
| Σ3(1 1 $\bar{2}$) | 130.2 | 7.1 | 2.9 | 156 | Σ41(4 $\bar{4}$ 3) | 157.1 | 2.9 | 18.5 | 490 |
| Σ5(0 1 3) | 117.5 | 4.1 | 6.5 | 180 | Σ41(0 1 9) | 147.6 | 4.1 | 18.5 | 652 |
| Σ5(0 1 3) | 129.2 | 4.1 | 9.1 | 280 | Σ41(0 5 4) | 156.9 | 4.1 | 26.2 | 980 |
| Σ7(2 $\bar{3}$ 1) | 138.5 | 7.1 | 13.3 | 756 | Σ43(3 $\bar{3}$ $\bar{5}$) | 161.0 | 2.9 | 19.0 | 516 |
| Σ9(1 $\bar{1}$ 4) | 139.1 | 2.9 | 12.3 | 288 | Σ43(5 $\bar{5}$ 6) | 149.0 | 2.9 | 26.9 | 677 |
| Σ9(2 $\bar{2}$ 1) | 148.3 | 2.9 | 8.7 | 216 | Σ51(1 $\bar{1}$ $\overline{10}$) | 122.9 | 2.9 | 29.1 | 603 |
| Σ11(1 $\bar{1}$ $\bar{3}$) | 136.2 | 2.9 | 9.6 | 220 | Σ51(5 $\bar{5}$ 1) | 176.1 | 2.9 | 20.6 | 612 |
| Σ11(3 $\bar{3}$ 2) | 134.9 | 2.9 | 13.6 | 308 | Σ53(0 9 5) | 126.5 | 4.1 | 21.0 | 632 |
| Σ13(3 $\bar{4}$ 1) | 125.2 | 7.1 | 18.1 | 930 | Σ57(2 $\bar{2}$ $\bar{7}$) | 184.7 | 2.9 | 21.9 | 682 |
| Σ13(0 1 5) | 125.1 | 4.1 | 10.4 | 310 | Σ57(7 $\bar{7}$ 4) | 129.4 | 2.9 | 30.9 | 672 |
| Σ13(0 3 2) | 148.2 | 4.1 | 14.7 | 520 | Σ57(7 $\bar{8}$ 1) | 130.8 | 7.1 | 12.6 | 680 |
| Σ17(2 $\bar{2}$ $\bar{3}$) | 134.9 | 2.9 | 12.0 | 272 | Σ59(3 $\bar{3}$ $\overline{10}$) | 132.6 | 2.9 | 31.5 | 704 |
| Σ17(3 $\bar{3}$ 4) | 142.2 | 2.9 | 16.9 | 406 | Σ59(5 $\bar{5}$ 3) | 125.2 | 2.9 | 22.2 | 468 |
| Σ17(0 1 4) | 135.9 | 4.1 | 16.9 | 544 | Σ61(0 6 5) | 127.6 | 4.1 | 31.9 | 972 |
| Σ17(0 5 3) | 143.2 | 4.1 | 11.9 | 406 | Σ61(0 1 11) | 133.5 | 4.1 | 22.6 | 718 |
| Σ19(1 $\bar{1}$ $\bar{6}$) | 126.1 | 2.9 | 17.8 | 378 | Σ65(0 9 7) | 138.2 | 4.1 | 23.3 | 768 |
| Σ19(3 $\bar{3}$ 1) | 143.2 | 2.9 | 12.6 | 304 | Σ67(3 $\bar{3}$ $\bar{7}$) | 132.4 | 2.9 | 23.7 | 530 |
| Σ21(4 $\bar{5}$ 1) | 132.1 | 7.1 | 7.7 | 416 | Σ67(7 $\bar{7}$ 6) | 141.6 | 2.9 | 33.3 | 796 |
| Σ25(0 1 7) | 144.5 | 4.1 | 14.5 | 498 | Σ73(1 $\bar{1}$ $\overline{12}$) | 148.0 | 2.9 | 34.8 | 868 |
| Σ25(0 4 3) | 163.7 | 4.1 | 20.4 | 798 | Σ73(6 $\bar{6}$ 1) | 139.2 | 2.9 | 24.7 | 578 |
| Σ27(1 $\bar{1}$ $\bar{5}$) | 128.0 | 2.9 | 15.0 | 324 | Σ73(0 11 5) | 147.9 | 4.1 | 24.7 | 867 |
| Σ27(5 $\bar{5}$ 2) | 150.3 | 2.9 | 21.3 | 538 | Σ81(4 $\bar{4}$ $\bar{7}$) | 146.9 | 2.9 | 26.1 | 646 |
| Σ29(0 2 5) | 131.8 | 4.1 | 22.0 | 688 | Σ81(7 $\bar{7}$ 8) | 153.6 | 2.9 | 36.9 | 958 |
| Σ29(0 7 3) | 124.1 | 4.1 | 15.6 | 458 | Σ83(1 $\bar{1}$ $\bar{9}$) | 149.7 | 2.9 | 26.3 | 664 |
| Σ33(1 $\bar{1}$ $\bar{8}$) | 132.8 | 2.9 | 23.4 | 524 | Σ83(9 $\bar{9}$ 2) | 158.3 | 2.9 | 37.2 | 992 |
| Σ33(4 $\bar{4}$ 1) | 141.2 | 2.9 | 16.6 | 394 | Σ89(2 $\bar{2}$ $\bar{9}$) | 152.6 | 2.9 | 27.3 | 700 |
| Σ37(0 1 6) | 149.2 | 4.1 | 24.9 | 884 | Σ97(6 $\bar{6}$ 5) | 161.6 | 2.9 | 28.4 | 774 |
| Σ37(0 7 5) | 139.9 | 4.1 | 17.6 | 586 | Σ99(7 $\bar{7}$ 1) | 162.2 | 2.9 | 29.0 | 792 |

## SM.4 Additional Figures

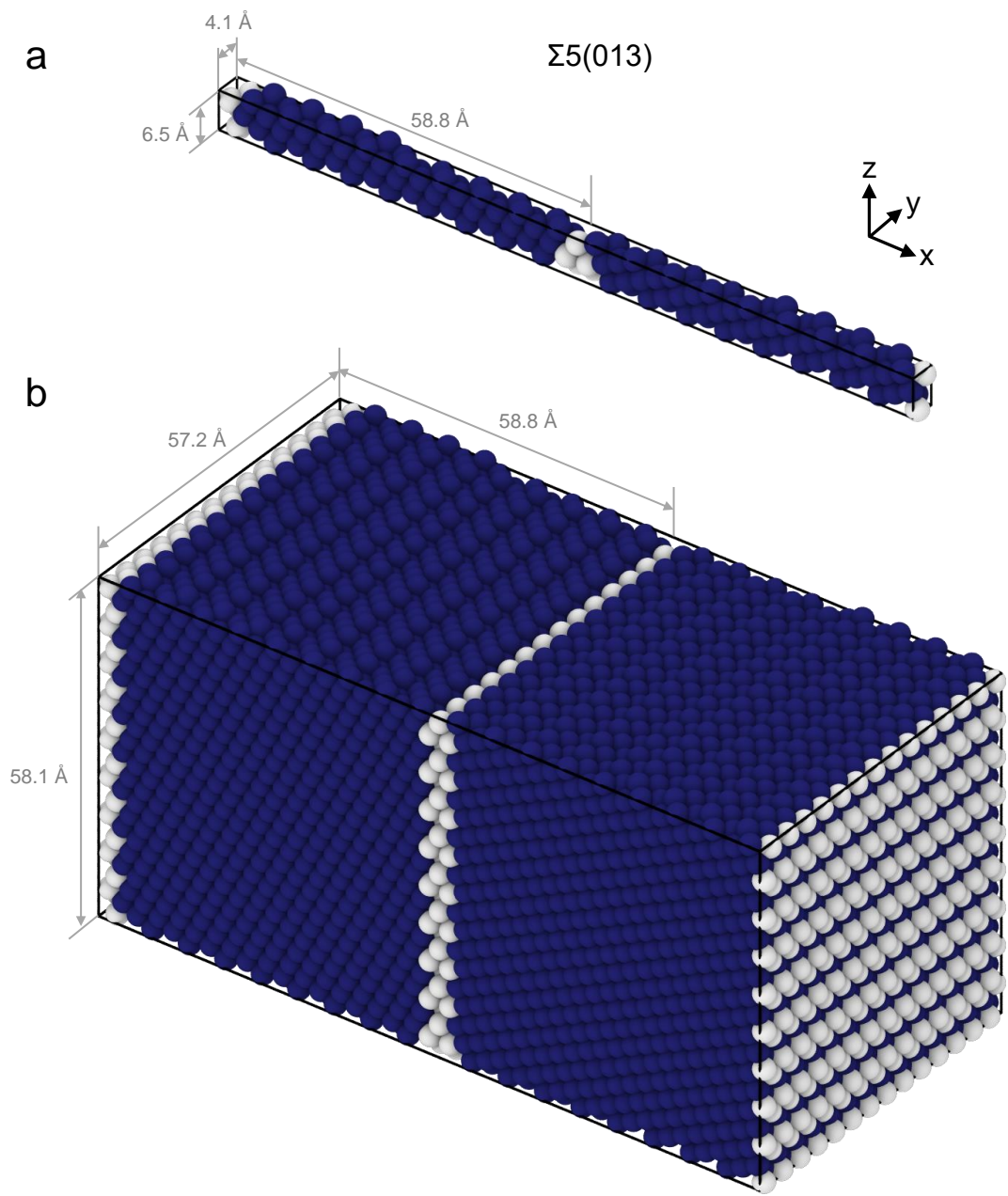

Fig. S1: An illustration of the Σ5(013) STGB: (a) The STGB structure as generated using the procedure detailed in section SM.3, and as listed in Table S1, and (b) the enlarged structure (with a size of >5 nm in the GB plane directions) used for segregation calculations to minimize self-interactions of the solute atom with its periodic image.

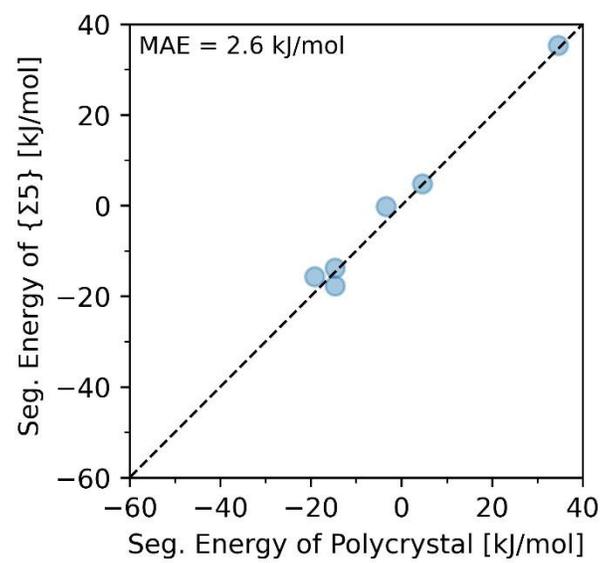

Fig. S2: A comparison of the segregation energies obtained by the seven unique sites of the {Σ5} STGBs against those of the closest polycrystalline sites in SOAP-descriptor space (i.e., environments most similar to the STGB sites).

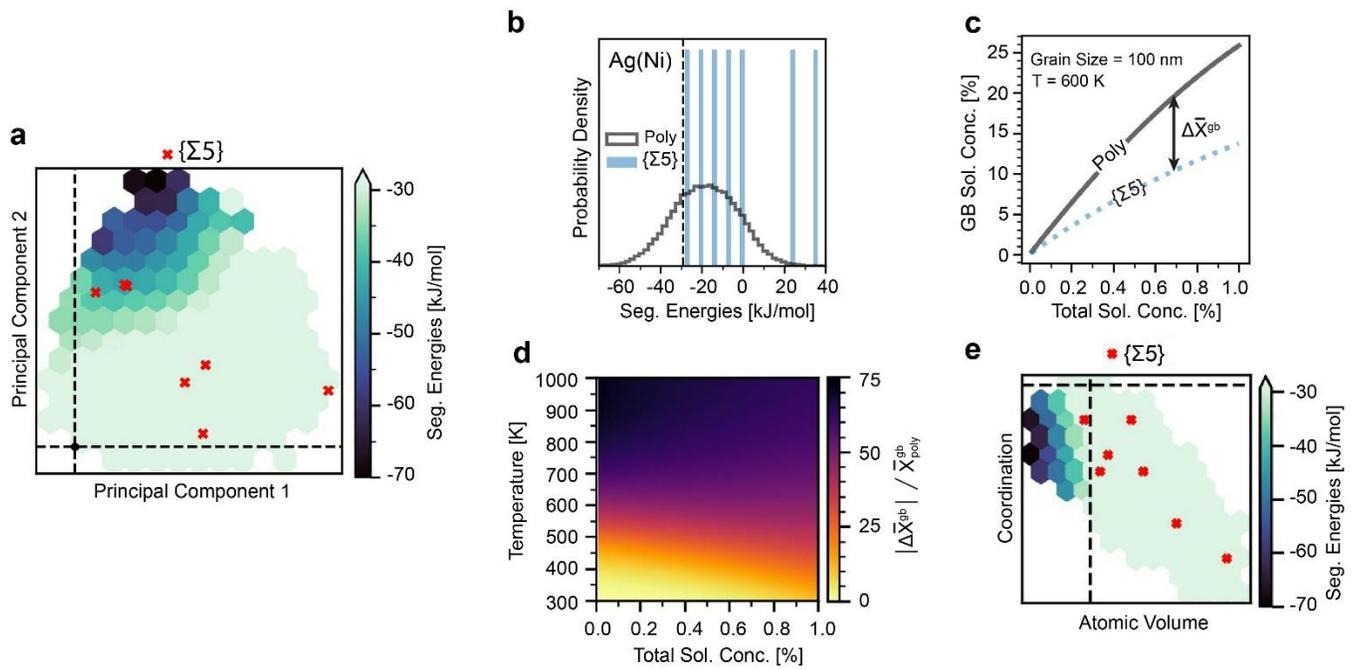

Fig. S3: A reproduction of the results of Fig. 2 in the manuscript using the more sophisticated embedded atom potential by Pan et al. [8] for Ag-Ni.

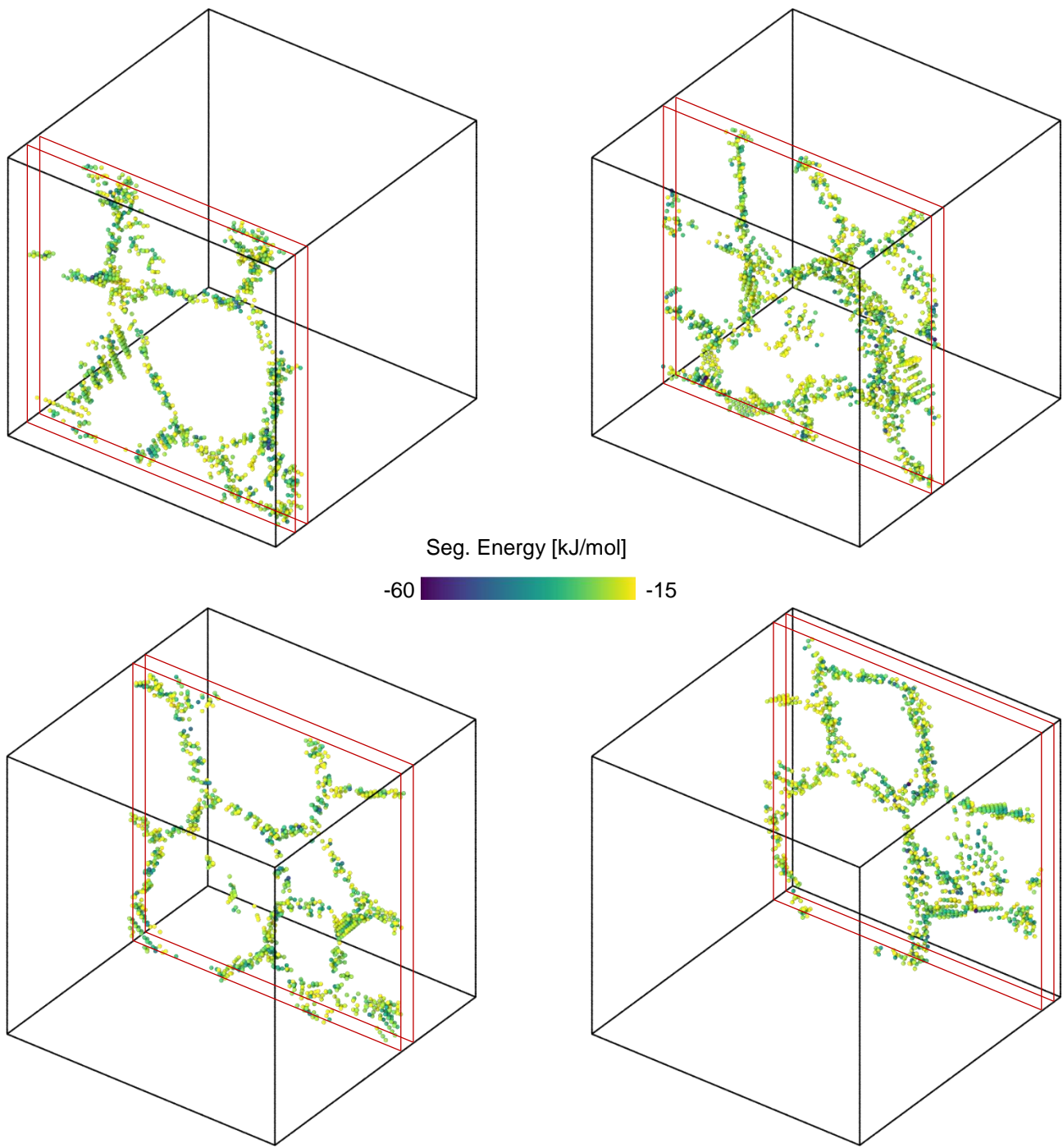

Fig. S4: Slices of the Ag polycrystal used in this work (Fig. 1(a) in the manuscript) that show the distribution of the top quartile (most attractive) GB sites for Ni segregation throughout the GB network.

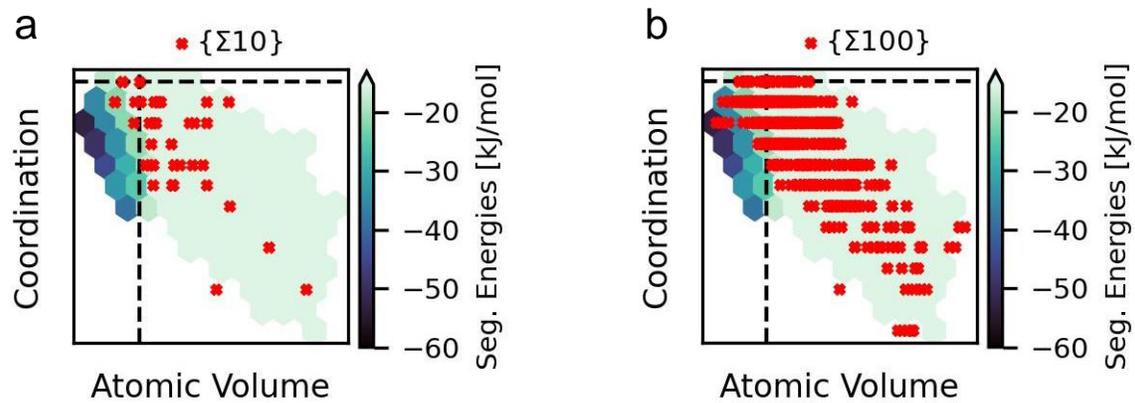

Fig. S5: The coverage of (a) {<Σ10} and (b) {<Σ100} STGB sets of the LAE space of the Ag polycrystal, as described by the classical descriptors of atomic volume and coordination.

## SM.5  Supplemental References